\documentclass[aps, prb, amsmath, amssymb, english, twocolumn,reprint,floatfix, superscriptaddress]{revtex4-2}

\usepackage[english]{babel}
\usepackage{graphicx}
\usepackage{verbatim}

\usepackage{epsfig}
\usepackage{epstopdf}
\usepackage{amsfonts}
\usepackage{amsthm}
\usepackage{amsmath}
\usepackage{amssymb}
\usepackage{color}
\usepackage[usenames,dvipsnames,svgnames,table]{xcolor}
\usepackage{hyperref} 
\hypersetup{pdfpagemode=FullScreen,colorlinks=true,linkcolor=NavyBlue,citecolor=BrickRed}
\usepackage{makeidx}
\usepackage{pifont}

\usepackage{booktabs}
\usepackage{xr}
\usepackage{color}
\usepackage{grffile}

\usepackage[ruled]{algorithm2e}
\usepackage[noend]{algpseudocode}
\usepackage{multirow}
\usepackage{type1cm} 
\usepackage{dcolumn}
\usepackage{bm}

\begin{document}

\title{Correlated electronic structure of high-temperature superconductor Ba$_2$CuO$_{3+\delta}$}
\author{Jing-Xuan Wang}\affiliation{School of Physics and Beijing Key Laboratory of Opto-electronic Functional Materials $\&$ Micro-nano Devices, Renmin University of China, Beijing 100872, China}\affiliation{Key Laboratory of Quantum State Construction and Manipulation (Ministry of Education), Renmin University of China, Beijing 100872, China}
\author{Rong-Qiang He}\email{rqhe@ruc.edu.cn}\affiliation{School of Physics and Beijing Key Laboratory of Opto-electronic Functional Materials $\&$ Micro-nano Devices, Renmin University of China, Beijing 100872, China}\affiliation{Key Laboratory of Quantum State Construction and Manipulation (Ministry of Education), Renmin University of China, Beijing 100872, China}
\author{Zhong-Yi Lu}\email{zlu@ruc.edu.cn}\affiliation{School of Physics and Beijing Key Laboratory of Opto-electronic Functional Materials $\&$ Micro-nano Devices, Renmin University of China, Beijing 100872, China}\affiliation{Key Laboratory of Quantum State Construction and Manipulation (Ministry of Education), Renmin University of China, Beijing 100872, China}\affiliation{Hefei National Laboratory, Hefei 230088, China}

\date{\today}

\begin{abstract}
Cuprate superconductors have attracted extensive attention due to high critical temperatures. Conventional cuprates typically contain perfect CuO$_2$ planes which are considered as a key factor to superconductivity since the superconductivity takes place in them. However, in Ba$_2$CuO$_{3+\delta}$ with $\delta=0.2$ and O-depleted CuO$_2$ planes, superconductivity still arises even with a transition temperature as high as 73 K. Using combined density functional theory and dynamical mean-field theory (DFT+DMFT) calculations, we investigated the electronic correlation and electronic structure of Ba$_2$CuO$_{3.25}$ with alternating quasi-one-dimensional (1D) CuO planes and O-depleted CuO$_2$ planes. We find that although different from the usual cuprates, the Cu atoms are still dominated by a 3$d^9$ configuration and the system is of a new kind of correlated single-orbital physics. The quasi-1D CuO planes, composed of parallel Cu-O chains, are slightly hole-doped quasi-1D Mott insulator, while the O-depleted CuO$_2$ planes are more hole doped, with a 2D correlated electronic structure, and may host superconductivity.

\end{abstract}


\maketitle

\section{INTRODUCTION}
In the past three decades, significant efforts have been devoted to discovering high critical temperature ($T_c$) superconductors in cuprate materials and the understanding of the superconducting mechanism~\cite{bednorz1986possible,wu1987PRL,keimer2015quantum,norman2016materials}. It is widely accepted that a complete CuO$_2$ plane is one of the key factors underlying high-temperature superconductivity~\cite{leggett2006we,dagotto1994correlated}. In traditional cuprate superconductors, such as the K$_2$NiF$_4$-type La$_{2-x}$Sr$_x$CuO$_4$, the Cu-O bond length between Cu and the apical O atoms is longer than that within the CuO$_2$ planes, forming a stretched octahedral structure. In this elongated octahedral environment, the energy of the $d_{x^2-y^2}$ orbital is higher than that of the $d_{z^2}$ orbital. Therefore, the $d_{z^2}$ orbital is fully occupied, while the $d_{x^2-y^2}$ orbital serves as the primary active channel for the doped carriers, which enhances in-plane electron correlations and promotes Cooper pair formation~\cite{aohta1991prb,peng2017np,chen1992prl,Sakakibara2020prl}. Previous studies have indicated that the elongation of the octahedra along the $c$-axis enhances the splitting between the $d_{x^2-y^2}$ and $d_{z^2}$ orbital energy levels, resulting in a higher superconducting critical temperature~\cite{peng2017np,botana2020prx}.

The successfully synthesized Ba$_2$CuO$_{3+\delta}$ compound under high pressure and high temperature exhibits superconductivity with a transition temperature above 70~K, more than twice that of bulk La$_{2-x}$Sr$_x$CuO$_4$~\cite{Li2019superconductivity}. This material does not conform to the conventional understanding of the elongated octahedral structure and exhibits distinct crystal-field and electronic structure. It appears with some different characteristics: in Ba$_2$CuO$_{3+\delta}$, the distance between Cu and the apical O atoms is shorter than that within the CuO$_2$ plane, resulting in the compression of the CuO$_6$ octahedral structure. This compression triggers the reversal of orbital energy levels, making the energy of the $d_{z^2}$ orbital higher than that of the $d_{x^2-y^2}$ orbital. In addition, the current consensus is that the optimal hole or electron doping concentration for the $T_c$ dome of cuprate superconductors is 0.15, while the hole-doping concentration in Ba$_2$CuO$_{3+\delta}$ is as high as 0.4 when the O atom defect $\delta = 0.2$, far exceeding the traditional optimal doping level, but still maintains strong superconductivity~\cite{keimer2015quantum,djscala2019pnas}. The unusual features of compressed octahedra and overdoping provide opportunities to exploring new pairing mechanisms in cuprate superconductors.

For Ba$_2$CuO$_{3+\delta}$, theoretical studies do not converge to date. There are works suggesting that a quasi-1D $d_{b^2-c^2}$ band is prone to strong antiferromagnetic fluctuations, thereby favoring superconductivity~\cite{liu2019prm,worm2022prb}. The pairing symmetry may involve an admixture of $s$- and $d$-wave components~\cite{gao2021nodeless,zegrodnik2021superconducting}. In the overdoped regime, an extended $s$-wave superconducting dome appears, in which the Cu $d_{z^2}$ orbital plays an active role in the pairing interaction~\cite{jiang2021prb,maier2019prb}. Similarly, calculations on the Lieb lattice model suggested a close competition between $s$- and $d$-wave pairings, with possible coexistence of intra- and inter-orbital pairing channels~\cite{Yamazaki2020prr}. Experimental investigations, including angle-resolved photoemission spectroscopy (ARPES) and resonant inelastic X-ray scattering (RIXS), have revealed enhanced nearest-neighbor attractive interactions (presumably originating from electron-phonon coupling) in doped quasi-1D cuprates in Ba$_2$CuO$_{3+\delta}$~\cite{chen2021science,li2025prl}. Despite these advances, there is still controversy in understanding the mechanism of high-temperature superconductivity in Ba$_2$CuO$_{3+\delta}$, particularly regarding the interplay between pairing symmetry and the contribution of different orbitals.

In this paper, we present the correlation properties and electronic structure of Ba$_2$CuO$_{3+\delta}$ through DFT and DFT+DMFT calculations. The most energetically stable bilayer crystal structure [Fig.~\ref{fig:structure}(b)] proposed in Ref.~\cite{liu2019prm,worm2022prb} are adopted. To achieve better orbital resolution, we introduce new local basis vectors for the Cu 3$d$ orbitals. Through comparative study, we find that the electronic correlations of the Cu(o) $d_{a^2-c^2}$ and Cu(p) $d_{b^2-c^2}$ orbitals in Ba$_2$CuO$_{3.25}$ are weaker than Cu $d_{b^2-c^2}$ in Ba$_2$CuO$_{3}$. This reduced correlation is primarily due to hole doping at both the Cu(o) and Cu(p) atoms, with the Cu(o) $d_{a^2-c^2}$ orbital exhibiting more hole doping. Moreover, our analysis of the band structure shows that the strongly correlated Cu(o) $d_{a^2-c^2}$ orbital in layer 1 possesses a pronounced two-dimensional (2D) character, in contrast to the quasi-1D nature of the Cu(p) in layer 2. One possible scenario is that the more hole doping and the 2D character of the Cu(o) $d_{a^2-c^2}$ orbital promote the emergence of superconductivity. These results motivate us to consider the role of enhanced hole doping and orbital dimensionality in facilitating superconductivity, offering new insights into the interplay between electronic correlations, orbital selectivity, and high-$T_c$ superconductivity in cuprates.

\section{METHODS}
In this study, the DFT plus DMFT calculations with the full charge-self consistency were performed using EDMFT package~\cite{haule2010dynamical,haule2015free}, based on WIEN2K~\cite{Blaha2020jcp}. Before performing DFT+DMFT, we first used the Vienna $ab$ $initio$ simulation package (VASP) software for structural optimization and electronic structure calculations~\cite{kresse1994norm,kresse1996efficiency}. We adopted the generalized gradient approximation (GGA) and the Perdew-Burke-Ernzerhof (PBE) functions to describe exchange-correlation energy~\cite{perdew1996generalized}. The muffin tin radii ($R_{MT}$) for Ba, Cu, and O were 2.48, 1.84, and 1.68 bohr, respectively. The maximum modulus for the reciprocal vectors $K_{\max}$ was chosen such that $R_{MT} \times K_{\max}=7.0$. A 7$\times$7$\times$7 $k$ mesh was used in the WIEN2K calculation and a 64$\times$64$\times$64 $k$ mesh was used in calculating the spectral function. Besides, the Hubbard $U$ and Hund’s coupling $J_H$ were chosen to be 7.5 and 1.0 eV, respectively, which are typically effective values for Cu 3$d$ electrons~\cite{dudarev1998electron,anisimov1991band}. We used the continuous time quantum Monte Carlo solver to solve the DMFT mapped quantum impurity problem~\cite{gull2011continuous}. The number of the Monte Carlo steps was set to 3$\times10^7$. The exact double-counting scheme was used for the self-energy function~\cite{haule2015exact}. Since there are three nonequivalent Wyckoff positions of Cu atoms in Ba$_2$CuO$_{3.25}$, we considered three impurity problems in DMFT. A single impurity problem was considered for Ba$_2$CuO$_3$. Since the $t_{2g}$ orbitals are fully filled, only the two $e_g$ orbitals were considered to be correlated. The self-energy functions at real frequencies were obtained by the analytic continuation method of maximum entropy. The mass enhancement was calculated with ${m^* / m}=1 / Z=1-\left.\left(\partial \operatorname{Re} \sum(\omega) / \partial \omega\right)\right|_{\omega=0}$~\cite{gull2011continuous}. We considered only the paramagnetic state and neglected the spin-orbital coupling effects.

\section{RESULTS}

\begin{figure}[htb]
  \centering
  \includegraphics[width=8.6cm]{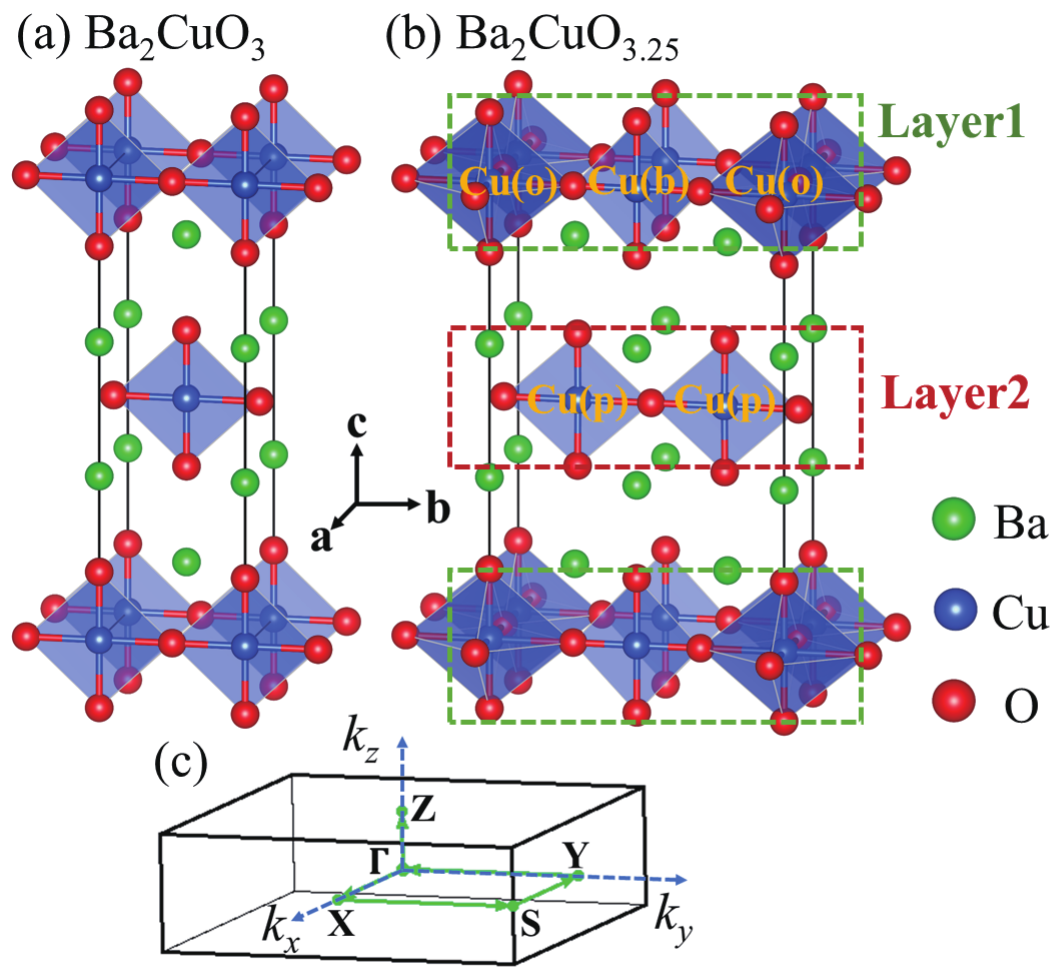}
  \caption{The crystal structure of (a) Ba$_2$CuO$_3$, featuring one-dimensional Cu-O chains, and (b) Ba$_2$CuO$_{3.25}$, featuring Cu-O bilayers. (c) shows the high-symmetry path in the first Brillouin zone. The high-symmetry path from $\rm Y$ to $\Gamma$ lies along $b$, and thus probes the dispersion along the chain. The bilayer Ba$_2$CuO$_{3.25}$ structure consists of alternating two-dimensional planes and quasi-one-dimensional Cu-O chains, denoted as layers 1 and 2, respectively. In layer 1, six-coordinated Cu atoms are denoted as Cu(o). The Cu(o)O$_6$ octahedra are connected directly in direction $a$ and are connected by Cu(b)O$_4$ squares as bridges in direction $b$. In layer 2, four-coordinated Cu atoms are denoted as Cu(p).}
  \label{fig:structure}
\end{figure}

\subsection{Crystal structures}

Experiments and theoretical simulations have revealed that oxygen vacancies in Ba$_2$CuO$_{4-y}$ predominantly occur within the CuO$_2$ planes rather than at apical oxygen sites~\cite{Li2019superconductivity,liu2019prm}. As shown in Fig.~\ref{fig:structure}, the crystal structure of Ba$_2$CuO$_3$ is composed of 1D Cu-O chains, where each Cu atom is four-coordinated with four O atoms nearby, forming a planar square. For the bilayer Ba$_2$CuO$_{3.25}$, the crystal structure features an alternating stacking of octahedral layers (layer 1) and 1D chain layers (layer 2). This structure has the lowest energy~\cite{liu2019prm,twoband2021prb}. Within layer 1, CuO$_6$ octahedra are interconnected by Cu-O square units that serve as bridging motifs. After structural relaxation, the Cu(o)-O bond lengths are 1.941, 2.398, and 1.994~\AA\ along the $a$, $b$, and $c$ axis, respectively, resulting in an elongated octahedron along the $b$-axis. To better describe electronic structure, we redefine the local basis vectors as follows: for Cu(p) and Cu(b) sites, the $bc$ plane is taken as the $xy$ plane, so that the $d_{x^2-y^2}$ orbital becomes the $d_{b^2-c^2}$ orbital in this new basis; for Cu(o), the $ac$ plane is taken as the $xy$ plane, transforming the $d_{x^2-y^2}$ orbital into the $d_{a^2-c^2}$ orbital.

The CuO$_6$ octahedron is stretched along the $b$ direction so that the $d_{b^2}$ orbital has a low energy, which is similar to the $d_{a^2}$ orbital of the Cu(p) in layer 2 and the Cu(b) in layer 1 without the apical O. As a result, both the $d_{b^2}$ of Cu(o) and the $d_{a^2}$ of Cu(p) and Cu(b) possess low energies and tend to be fully filled.

\begin{table}[t]
  \centering
  \renewcommand\arraystretch{1.5}
  \caption{Local orbital occupation number $N_d$ and effective mass enhancement $m^*/m$ of Cu atoms in Ba$_2$CuO$_{3+\delta}$ respectively with $\delta=0$ and $0.25$. The orbitals $d_{z^2}$ and $d_{x^2-y^2}$ are defined with respect to local basis vectors.}
  \label{tab1}
  \begin{tabular*}{8.6cm}{@{\extracolsep{\fill}} lcccc}
    \hline\hline
    \multirow{2}{*}{\centering Ba$_2$CuO$_{3+\delta}$} & 
    \multicolumn{3}{c}{$N_d$} & 
    \multicolumn{1}{c}{$m^*/m$} \\
    \cmidrule(lr){2-4} \cmidrule(lr){5-5}
    & $d_{z^2}$ & $d_{x^2-y^2}$ & $e_g$ & $d_{x^2-y^2}$ \\
    \hline
    $\delta=0$ Cu        & 1.793 & 1.355 & 3.148 & 2.84 \\
    $\delta=0.25$ Cu(p)     & 1.807 & 1.264 & 3.071 & 2.22 \\
    $\delta=0.25$ Cu(o)     & 1.850 & 1.181 & 3.031 & 2.47 \\
    $\delta=0.25$ Cu(b)     & 1.817 & 1.120 & 2.937 & --   \\
    \hline\hline
  \end{tabular*}
\end{table}

\begin{table}[t]
  \centering
  \renewcommand\arraystretch{1.5}
  \caption{The weights (\%) of the Cu-$e_g$ orbital local multiplets for Ba$_2$CuO$_{3+\delta}$ respectively with $\delta=0$ and $0.25$ calculated by DFT+DMFT at 290 K. The good quantum numbers $N_\Gamma$ and $S_z$ denote the total occupancy and total spin of the Cu-$e_g$ orbital, which are used to label different local spin states. The effective local spin $S_{\rm eff}$ is calculated through $S_{\rm eff}(S_{\rm eff}+1)=3\langle S_{z}^2 \rangle$.}
  \label{tab2}
  \begin{tabular*}{8.6cm}{@{\extracolsep{\fill}} lccccccc}
    \hline\hline
    $N_\Gamma$& 0 & 1 & 2 & 2 &3 & 4&$ $\\
    $S_z$& 0 & 1/2 & 0 & 1 &1/2 & 0&$S_{\rm eff}$\\
    \hline
    $\delta=0$ Cu& 0.000& 0.002 & 0.057 & 0.041 & 0.659 & 0.240&0.433\\
    $\delta=0.25$ Cu(p)& 0.000 & 0.004 & 0.080 & 0.046 & 0.664 & 0.205&0.443\\
    $\delta=0.25$ Cu(o)& 0.000 & 0.003 & 0.092 & 0.038 & 0.700 & 0.168&0.443\\
    $\delta=0.25$ Cu(b)&0.000 & 0.011 & 0.151 & 0.052 & 0.625 & 0.162&0.440\\
    \hline\hline
  \end{tabular*}
\end{table}

\subsection{Local occupancies}

Doping is a key factor to unconventional superconductivity. Here, we calculated the local orbital occupation numbers $N_d$ by DFT+DMFT at 290~K to analyze possible doping through orbital occupancies. The results are shown in Table~\ref{tab1}. The $d_{z^2}$ and $d_{x^2-y^2}$ orbitals are defined with respect to local basis vectors. The local $e_g$ orbital occupancies of both Cu(p) and Cu(b) are approximately 3.0, consistent with a nominal 3$d^9$ configuration. The Cu(o) atoms here and Cu atoms in Ba$_2$CuO$_4$ have similar coordination environments and are expected to have a 3$d^7$ configuration. However, the occupation number of Cu(o) $e_g$ orbitals is close to 3, which deviates from the nominal 3$d^7$ configuration. This deviation arises from charge transfer between Cu and O atoms. Although the number of electrons on the $p$ orbital of O atom cannot be determined precisely, we infer that the O atoms coordinated to Cu(o) exhibit a significant deviation from a closed-shell configuration.


For all Cu-$d_{z^2}$ orbitals, the local occupancies are close to 2, indicating that these orbitals are fully occupied. There are 1.355, 1.264 and 1.181 electrons residing in the Cu $d_{b^2-c^2}$ in Ba$_2$CuO$_{3}$, Cu(p) $d_{b^2-c^2}$ and Cu(o) $d_{a^2-c^2}$ orbitals in Ba$_2$CuO$_{3.25}$, respectively. As shown in Table~\ref{tab2}, the $e_g$ orbitals of all the three kinds of Cu atoms are dominated by the configuration of $N_\Gamma=3$ and $S_z=1/2$, with an effective spin of about 0.44, which is about 88$\%$ of a spin $\frac{1}{2}$. These findings indicate that Ba$_2$CuO$_{3.25}$ is a single-orbital hole-doped superconductor. For Ba$_2$CuO$_{3.25}$, the $d_{b^2-c^2}$ orbital occupancies of the four-coordinated Cu(p), compared with Ba$_2$CuO$_{3}$, is reduced, indicating inter-layer charge transfer and hence that Cu(p) is slightly hole doped. Furthermore, the $d_{a^2-c^2}$ orbital occupancies of Cu(o) is even lower, suggesting more hole doping in these orbitals.

\begin{figure}[htb]
  \centering
  \includegraphics[width=8.6cm]{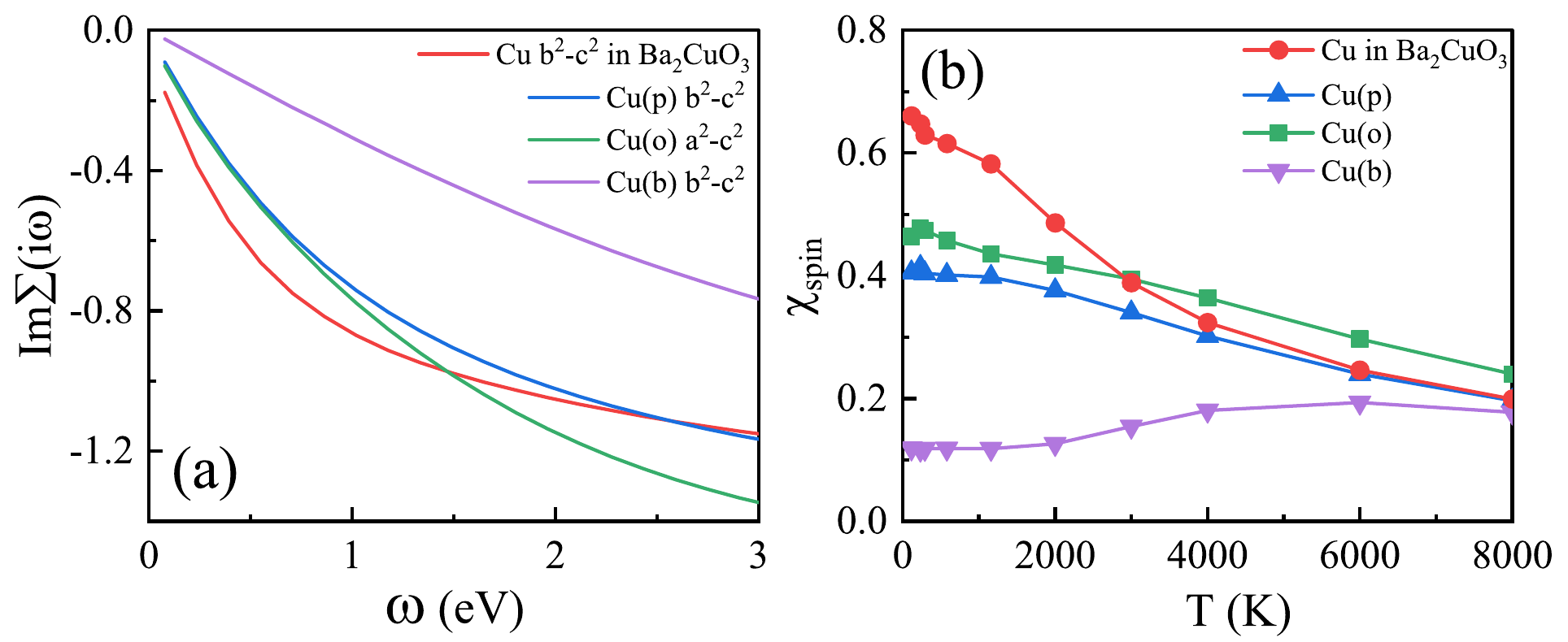}
  \caption{(a) Imaginary parts of the self-energy functions at the Matsubara axis $\text{Im}\Sigma(i\omega)$ at 290 K for Cu atoms in Ba$_2$CuO$_{3+\delta}$. (b) Static local spin susceptibilities $\chi_{\mathrm{spin}}$ for Cu atoms in Ba$_2$CuO$_{3+\delta}$ as functions of temperature.}
  \label{fig:self-energy}
\end{figure}

\subsection{Self-energies}

To explore the electronic correlation, we show the imaginary parts of the Matsubara self-energy $\text{Im}\Sigma(i\omega)$ for the Cu atoms in Ba$_2$CuO$_3$ and Ba$_2$CuO$_{3.25}$ in Fig.~\ref{fig:self-energy}(a). For Ba$_2$CuO$_3$, the magnitudes of $\text{Im}\Sigma(i\omega)$ for $d_{b^2-c^2}$ orbital have relatively large value at low frequencies,  which exhibits strong electronic correlation. As is well known, Ba$_2$CuO$_{3}$ is an antiferromagnetic Mott insulator where Cu atoms have a 3$d^9$ configuration with half-filled $d_{b^2-c^2}$ orbitals. However, the Mott insulating behavior of the $d_{b^2-c^2}$ state is not observed, because the antiferromagnetic order was not taken into account. For Ba$_2$CuO$_{3.25}$, one can see that $\text{Im}\Sigma(i\omega)$ for the Cu(o) $d_{a^2-c^2}$ and Cu(p) $d_{b^2-c^2}$ orbitals are very close and deviate from linearity at low frequencies, which exhibit non-Fermi liquid behavior. The magnitudes of $\text{Im}\Sigma(i\omega)$ at low frequencies for the Cu(p) $d_{b^2-c^2}$ orbital in Ba$_2$CuO$_{3.25}$ are smaller than those in Ba$_2$CuO$_3$, implying that for Cu atoms with identical coordination environments, the oxygen concentration in the system controls the level of electronic doping, which in turn affects the electronic correlation. Moreover, the $d_{b^2-c^2}$ orbital of Cu(b) atom exhibit a weakly correlated band insulating states with a small self-energy.

\subsection{Effective mass enhancement}
Then, we consider the effective mass enhancement $m^*/m$ at 290~K. Ba$_2$CuO$_{3}$ exhibits strong correlation, with an effective mass $m^*/m$ $\sim$ 2.84. In Ba$_2$CuO$_{3.25}$, both the Cu(p) $d_{b^2-c^2}$ and Cu(o) $d_{a^2-c^2}$ orbitals also exhibit strong correlation, with effective masses of 2.22 and 2.47, respectively, although their correlation is slightly weaker than that in Ba$_2$CuO$_{3}$. The weaker electronic correlation may be explained by the hole doping at the Cu sites in Ba$_2$CuO$_{3.25}$, in comparison to Ba$_2$CuO$_{3}$. 

\subsection{Static local spin susceptibilities}
Furthermore, we calculated the temperature dependence of the static local spin susceptibilities, $\chi_{\rm spin}$ $=$ $\int_{0}^{\beta}$$\langle{S_z(\tau)S_z(0)} \rangle$$d\tau$, for Ba$_2$CuO$_{3+\delta}$ with $\delta=0$ and 0.25, as shown in Fig.~\ref{fig:self-energy}(b). For Ba$_2$CuO$_{3}$, the local magnetic susceptibility increases steadily with decreasing temperature, indicating the presence of unquenched localized magnetic moments and strong electronic correlation. For Ba$_2$CuO$_{3.25}$, the local magnetic susceptibilities of both Cu(p) and Cu(o) also exhibit a similar increasing trend as the temperature decreases. As the temperature is further reduced to 145 K, the susceptibility slightly decreases and saturates. These results indicate that the local susceptibility approaches a constant value at low temperatures, suggesting that the spin degrees of freedom are screened~\cite{deng2019signatures}. The local magnetic susceptibility of Cu(b) exhibits a non-monotonic temperature dependence, deviating from the Curie law. The Cu(b) local magnetic susceptibility is relatively small, which indicated that the Cu(b) may be in an insulating state with a narrow band gap~\cite{addendum2000nature,krannich2015magnetic}.

\begin{figure}[htb]
  \centering
  \includegraphics[width=8.6cm]{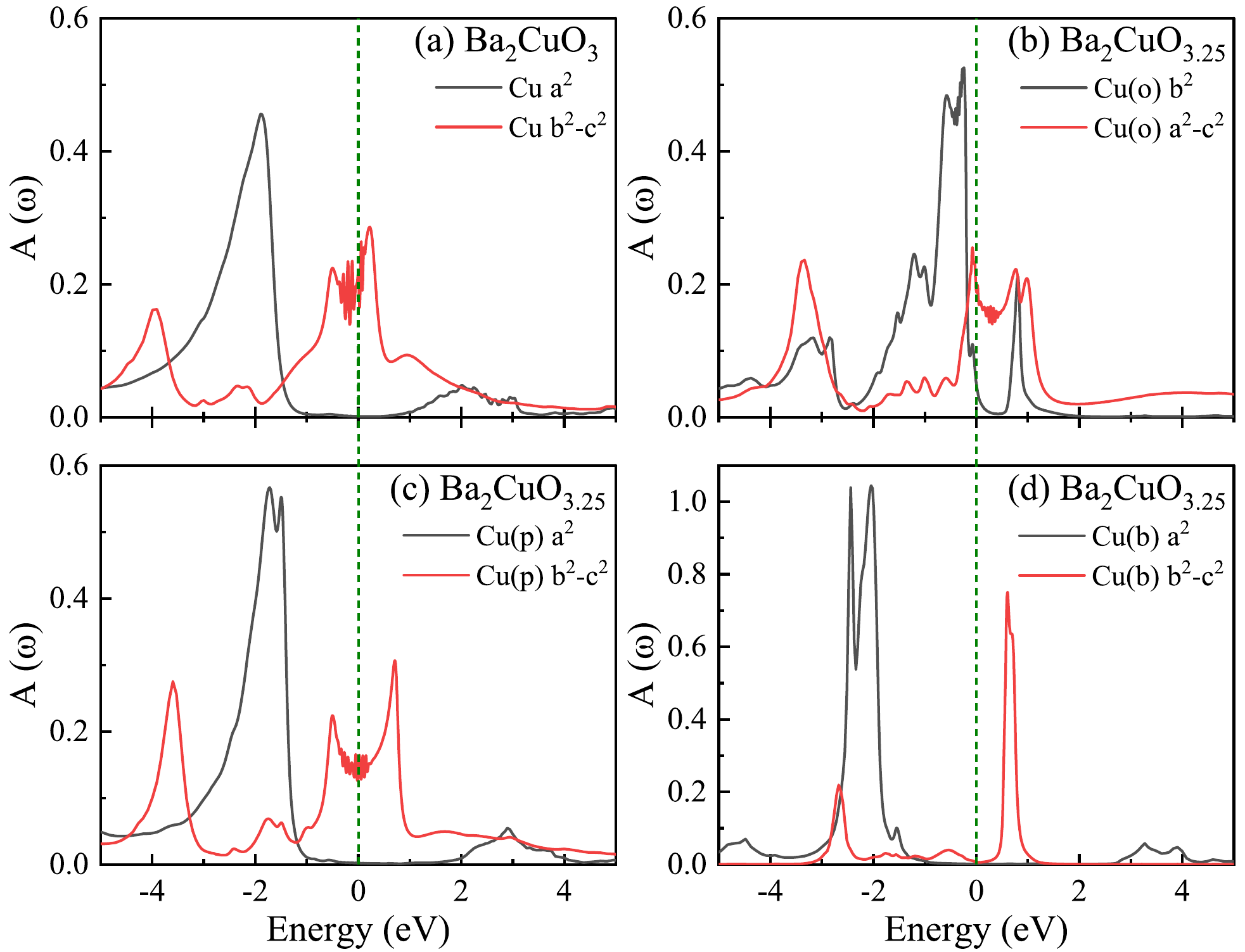}
  \caption{Calculated spectral functions $A(\omega)$ for the Cu-3$d$ orbitals in (a) Ba$_2$CuO$_3$, and (b-d) Ba$_2$CuO$_{3.25}$ at 290 K by using DFT+DMFT. The vertical dashed lines denote the Fermi energy level.}
  \label{fig:dos}
\end{figure}

\begin{figure*}[htb]
  \centering
  \includegraphics[width=17.2cm]{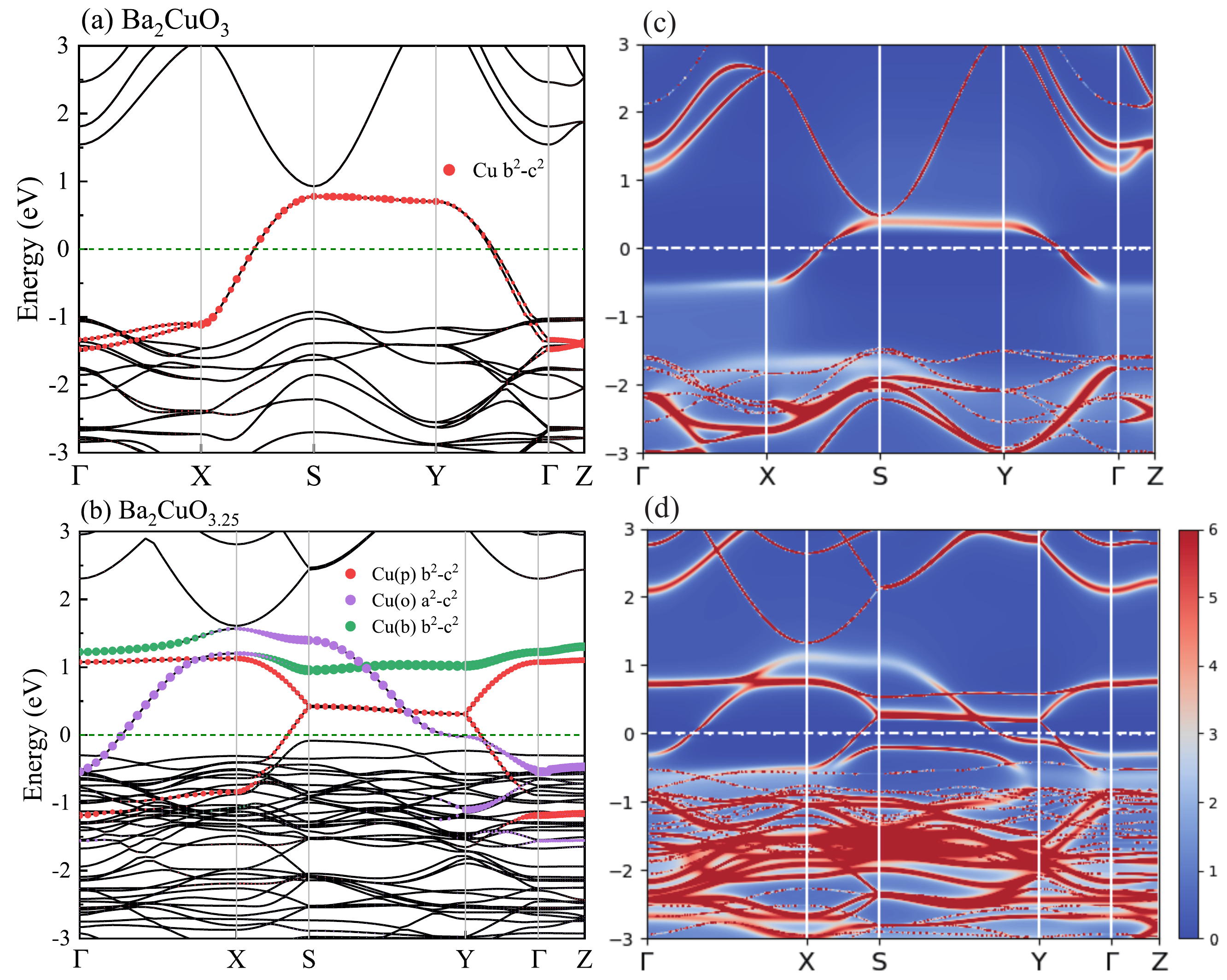}
  \caption{Band structures of Ba$_2$CuO$_{3}$ and Ba$_2$CuO$_{3.25}$ obtained by (a, b) DFT and (c, d) DFT+DMFT calculations at 290~K. The dashed lines at 0 eV indicate the Fermi level.}
  \label{fig:band-dft}
\end{figure*}

\subsection{Local spectral functions}
The local spectral functions $A(\omega)$ of the Cu $e_g$ electrons in Ba$_2$CuO$_{3+\delta}$ calculated with the DFT+DMFT approach at 290~K are presented in Fig.~\ref{fig:dos}. For Ba$_2$CuO$_3$, the spectral function of the $d_{a^2}$ orbital exhibits a band gap of approximately 2 eV, indicating an insulating state. In contrast, the spectral function of the $d_{b^2-c^2}$ orbital displays two peaks at energies of approximately $-$0.5 eV and 0.5 eV, forming a pseudogap. Without considering antiferromagnetic order, the system fails to realize a Mott insulating state and remains a correlated metal.

The spectral functions of the $d_{a^2}$ and $d_{b^2-c^2}$ orbitals of Cu atoms in Ba$_2$CuO$_{3}$ and Cu(p) atoms in Ba$_2$CuO$_{3.25}$ are shown in Figs.~\ref{fig:dos}(a) and (c), respectively. We note that the spectral functions of the Cu(p) $d_{b^2-c^2}$ orbital in Ba$_2$CuO$_{3.25}$ is similar to that of Cu atom in Ba$_2$CuO$_3$. The peaks of the Cu $d_{b^2-c^2}$ orbital appear at about $-$4.0 eV, $-$2.2 eV, and $-$0.5 eV in Ba$_2$CuO$_3$, while at about $-$3.8 eV, $-$1.8 eV, and $-$0.5 eV in Ba$_2$CuO$_{3.25}$. This indicates that the Cu(p) atoms in Ba$_2$CuO$_{3.25}$ are doped with holes compared to Ba$_2$CuO$_3$, indicating that there is electron charge transfer from layer 2 to layer 1 in Ba$_2$CuO$_{3.25}$.

In Fig.~\ref{fig:dos}(b), the spectral function of the $d_{a^2-c^2}$ orbital of the Cu(o) atom exhibits metallic characteristics, with peak features also similar to those of the Cu(p) atom in Ba$_2$CuO$_{3.25}$. The $d_{a^2-c^2}$ orbital of Cu(o) shows a higher degree of hole doping compared to the $d_{b^2-c^2}$ orbital of Cu(p) in Ba$_2$CuO$_{3.25}$. Additionally, the spectral function of Cu(b) in Ba$_2$CuO$_{3.25}$ exhibits a small gap at the Fermi level, showing that it is in an insulating state. Notably, this insulating behavior is not driven by Coulomb repulsion but rather arises from its electronic band [see Figs.~\ref{fig:band-dft}(b) and (d)], indicative of a band insulator. The narrow peak near 0.6~eV implies flat band and localized states.

\subsection{Correlated band structures}

We now turn to discuss the band structures obtained with DFT and DFT+DMFT. Only paramagnetic states are considered. The band structures of Ba$_2$CuO$_{3}$ are shown in Figs.~\ref{fig:band-dft}(a) and (c). We find that Ba$_2$CuO$_3$ is a strongly correlated metal. After considering electronic correlation with DMFT, the bandwidth of the $d_{b^2-c^2}$ orbital in Ba$_2$CuO$_{3}$ decreases from approximately 2.2 eV to 0.8 eV near the Fermi level, which shows significant renormalization due to electronic correlation.

The band structures of Ba$_2$CuO$_{3.25}$ are shown in Figs.~\ref{fig:band-dft}(b) and (d). Both the Cu(p) $d_{b^2-c^2}$ and Cu(o) $d_{a^2-c^2}$ orbitals contribute to the electronic states near the Fermi level. The band renormalization in Ba$_2$CuO$_{3.25}$ is slightly weaker than that in Ba$_2$CuO$_{3}$. This indicates that the electronic correlation for the Cu(p) $d_{b^2-c^2}$ and Cu(o) $d_{a^2-c^2}$ orbitals in Ba$_2$CuO$_{3.25}$ is weaker than that for the Cu $d_{b^2-c^2}$ orbital in Ba$_2$CuO$_{3}$. The blurry spectra around $\pm 1$ eV, originating from the Cu(p) $d_{b^2-c^2}$ and Cu(o) $d_{a^2-c^2}$ orbitals in Ba$_2$CuO$_{3.25}$, exhibits electronic incoherence due to strong correlation. There is orbital hybridization between Cu(o) $d_{a^2-c^2}$ and Cu(b) $d_{b^2-c^2}$ near the Fermi level.

For all Cu atoms in Ba$_2$CuO$_{3+\delta}$, the $d_{z^2}$ band, defined with respect to their corresponding local basis, lie entirely below the Fermi level, and their energy positions can be identified from the density of states [see Fig.~\ref{fig:dos}]. For Ba$_2$CuO$_3$, the energy bands exhibit strong dispersion along the high-symmetry paths X $\to$ S and Y $\to$ $\Gamma$ (correspond to the chain direction), while keeping flat along $\Gamma$ $\to$ X and S $\to$ Y, indicating that Ba$_2$CuO$_3$ has a quasi-1D feature. Similarly, for Ba$_2$CuO$_{3.25}$, the Cu(p) in layer 2 exhibits an analogous feature, indicating that the layer 2 in Ba$_2$CuO$_{3.25}$ also displays an evident quasi-1D feature. For Cu(o), significant band dispersion is observed along both the S $\to$ Y and Y $\to$ $\Gamma$ paths, exhibiting a 2D character.

\section{Discussion and conclusion}

Previous studies have shown that reported quasi-1D superconductors, such as organic superconductors~\cite{jerome1980JPL,bechgaard1981prl}, chromium-based compounds ~\cite{superconductivity2015prx,mu2018prm,liu2023nature,chen2024unconventional}, bismuth sulfide~\cite{ma2024prb}, nickel-bismuth compounds~\cite{wang2018prb,Liu2023jssc} and so on, typically exhibit low superconducting transition temperatures, usually below 10~K. The emergence of superconductivity is associated with spontaneous symmetry breaking. In quasi-1D systems, strong fluctuations induced by low dimensionality, the difficulty in achieving long-range order, and Luttinger liquid behavior all contribute to the significant suppression of superconducting $T_c$. In contrast, quasi-2D Ba$_2$CuO$_{3+\delta}$~\cite{Li2019superconductivity} shows a superconducting transition temperature well above 10 K and no superconductivity was found in hole-doped quasi-1D Ba$_2$CuO$_3$~\cite{chen2021science,li2025prl}. Thus, we claim that the superconductivity of Ba$_2$CuO$_{3+\delta}$ is not driven by its quasi-1D electronic structure in layer 2 but the 2D electronic structure of Cu(o) $d_{a^2-c^2}$ orbital in layer 1. Several studies suggested that the pairing mechanism and superconducting carriers in Ba$_2$CuO$_{3+\delta}$ arise from double-orbital physics~\cite{zegrodnik2021superconducting,jiang2021prb,twoband2021prb}. However, we show that the system is still dominated by single-orbital physics, because all $d_{z^2}$ orbitals (in respective atomic local basis) are fully filled.

In conclusion, we have studied the electronic correlation and electronic structures of Ba$_2$CuO$_{3+\delta}$ respectively with $\delta=0$ and 0.25 by DFT+DMFT calculations. We find that these systems are still dominated by single-orbital physics since the $d_{z^2}$ orbital (in Cu-atom-dependent local basis) of each Cu atom is fully filled. In Ba$_2$CuO$_{3.25}$, both Cu(o) $d_{a^2-c^2}$ and Cu(p) $d_{b^2-c^2}$ orbitals exhibit strong electronic correlations, although weaker than the Cu $d_{b^2-c^2}$ orbital in Ba$_2$CuO$_{3}$. Analysis of the electron occupation numbers indicates that both the Cu(o) and Cu(p) sites in Ba$_2$CuO$_{3.25}$ are hole doped, with the Cu(o) site showing a higher level of hole doping compared to that in Ba$_2$CuO$_{3}$. Band structure calculations reveal that the Cu(p) atoms in layer 2 possess a quasi-1D character, similar to that observed in Ba$_2$CuO$_{3}$, while layer 1, characterized by strongly correlated Cu(o) $d_{a^2-c^2}$ orbital, exhibits a 2D feature. These findings provide further support for the importance of orbital selectivity and dimensionality in understanding superconductivity in cuprates. Overall, the strong correlations and 2D character of the Cu(o) $d_{a^2-c^2}$ orbital may be the key factors for the emergence of superconductivity in these materials.


\begin{acknowledgments}
This work was supported by the National Key R\&D Program of China (Grants No. 2024YFA1408601 and No. 2024YFA1408602) and the National Natural Science Foundation of China (Grant No. 12434009). J.X.W. was also supported by the Outstanding Innovative Talents Cultivation Funded Programs 2025 of Renmin University of China. Z.Y.L. was also supported by the Innovation Program for Quantum Science and Technology (Grant No. 2021ZD0302402). Computational resources were provided by the Physical Laboratory of High Performance Computing in Renmin University of China.
\end{acknowledgments}

\bibliography{lco}

\end{document}